# Rate coefficients of a roaming reaction H + MgH using the ring polymer molecular dynamics

Hui Yang,[a#] Wenbin Fan,[a#] Junhua Fang,[b] Jianing Song[a*] and Yongle Li[a,c*]

[a] *Department of Physics, International Center for Quantum and Molecular Structures, and Shanghai Key Laboratory of High Temperature Superconductors, Shanghai University, Shanghai 200444, China*

[b] *Department of Chemistry, Shanghai University, Shanghai 200444, China*

[c] *Devision of Chemistry and Chemical Engineering, California Institute of Technology, CA 91125*

The ring-polymer molecular dynamics (RPMD) was used to calculate the thermal rate coefficients of the multi-channel roaming reaction H + MgH $\rightarrow$ Mg + $H_2$. Two reaction channels, tight and roaming, are explicitly considered. This is a pioneering attempt of exerting RPMD method to multi-channel reactions. With the help of a newly developed optimization-interpolation protocol for preparing the initial structures and adaptive protocol for choosing the force constants, we have successfully obtained the thermal rate coefficients. The results are consistent with those from other theoretical methods, such as variational transition state theory (VTST) and quantum dynamics (QD). Especially, RPMD results exhibit negative temperature dependence, which is similar to results from VTST but different from ones from ground state QD calculations.

---

*Corresponding authors.

# These authors contributed equally.

E-mail address: js888@shu.edu.cn (JS) and yongleli@shu.edu.cn; yongle@caltech.edu (YL).

## I. INTRODUCTION

Recently, both experimental and theoretical studies have revealed that there exists a new reaction pathway called a "roaming mechanism" in the unimolecular dissociation of a closed-shell molecule [1-4]. In such kind of reactions, there are two reaction channels; one is the reaction system passes a traditional tight transition state (TS) from reactant state to the product mainly goes through the minimum energy path (MEP), the other is the system passes a loose TS, by forming a long lifetime complex with large interatomic distances [3]. Once the reaction system passes the roaming channel, the system will also visit a large region of the potential energy surface (PES), like herds of cattle roaming on the Great Plains. And that is where the name roaming comes from. Thus, it has been pointed out that vibrational and rotational distributions of the products obtained through the roaming channel are significantly different from those obtained via the traditional reaction of the products through traditional channels [4,5]. Till now, lots of reactions are found with the roaming channel. Among them, the H + MgH reaction is the simplest system containing roaming (There are three reaction channels of the title reaction. During the reaction, about 80% trajectories will follow the direct channel, leaving about 20% reaction trajectories go through the tight and roaming channels) [6]. And recently, with the a newly developed PES [6], Li et al. revealed that the H + MgH reaction mechanism in the tight and roaming channels is firstly an addition reaction forming a complex, and then dissociating to the product. Such a process would go through a tight or roaming transition state (t-TS, r-TS). Both tight and roaming channels are shown in Fig 1.

The rate coefficient provides an insight into the dynamic characteristics, since it is an experimental observable, and can also give a hint to the reaction mechanism. In the obtaining of rate coefficients, the theoretical method is indispensable, since it can cover a wider range of temperature, but also provide more dynamic details of the reaction. For reactions involving light atoms, such as the title reaction, the quantum effect is dominant so that quantum mechanical treatment is necessary. Although quantum dynamics calculations with wave packages provide a bottom-up method to calculate rate coefficients, the calculation cost is enormous for thermal rate coefficient calculations [7,8]. The most popular alternative is the transition state theory (TST) based on statistical ansatz purposed by Wigner, Erying, Evans, and Polanyi [9-11]. However, the results from TST are less reliable because TST neglects the recrossing effect and uses the semi-empirical method to estimate tunneling effects. It is also inaccurate when the reaction is in the deep tunneling region at low temperatures [9-12]. Although the VTST can minimize the recrossing within the TST framework, it still may be with large error [9-11]. Another alternative is the quasi-classical trajectory (QCT) method which can incorporate full recrossing since using classical dynamics to propagate the reaction trajectories but suffers other problems such as without quantum effects during propagation, and the ZPE leakage [8,13].

A newly devised method, ring polymer molecular dynamics (RPMD) [14-16] is proved to be a reliable alternative of QD during recent years. It is an approximate quantum mechanical method, based on the isomorphism between the statistical properties of the quantum system and a fictitious ring polymer consisting of classical beads connected by harmonic springs [17]. It shows a series of advantages. First of all, it can include ZPE and tunneling effect directly. At the high-temperature limit, the RPMD results converge to the classical limit [18]. And it is exact in the case of the harmonic barrier. It also has a well-defined short-time limit and provides an upper bound. It is also equivalent to the quantum transition-state theory in the limit of non-recrossing [18,19]. And RPMD results are also independent of the selection of the dividing surface [20,21].

So it's straightforward to consider investigating RPMD's validity of treating reactions with multi-channel reactions using the H + MgH reaction system, due to clear mechanism and high accurate PES. And specifically, for treating the multiple reaction channels, we have developed a protocol to obtain the initial configurations along both reaction channels for umbrella sampling. But considering the task of this work is illustrating calculation of multi-channel reactions using RPMD, and the only quantum dynamical result available for comparison is for the tight and roaming channels only, we focus on the tight and roaming channels in current work. The rest of this paper is organized as follows. Section 2 briefly describes the methodology. And the computational details summarized in Section 3. In Section 4, results obtained from this work are listed and discussed. The conclusion is reached in Section 5.

## II. COMPUTATIONAL STRATEGY

### A. Ring-polymer molecular dynamics rate theory

In this work, all RPMD calculations are carried out using the RPMDrate code [22]. Since details of RPMD have been well-reviewed elsewhere [14,15], we only give a brief summary here. We make use of the isomorphism between the quantum and corresponding classical model system, and define the ring-polymer Hamiltonian for the title reaction:

$$H_n\left(\vec{p},\vec{q}\right)=\sum_{i=1}^{3}\sum_{j=1}^{n}\left(\frac{\left|\vec{p}_i^{(j)}\right|^2}{2m_i}+\frac{m_i\omega_n^2}{2}\left|\vec{q}_i^{(j)}-\vec{q}_i^{(j-1)}\right|^2\right)+\sum_{j=1}^{n}V\left(\vec{q}_1^{(j)},\vec{q}_2^{(j)},\vec{q}_3^{(j)}\right), \quad (1)$$

where $\omega_n=\left(\beta_n\hbar\right)^{-1}$ is the force constant between two neighboring beads. Here $\beta_n=\beta/n$ with $\beta=1/\left(k_\mathrm{B}T\right)$ is the reciprocal temperature of the ring polymer system, $\vec{p}_i$, $\vec{q}_i$ and $m_i$ are the momentum, position, and mass of the $i$th atom, respectively. The potential energy $V(\vec{q}_1,\vec{q}_2,\vec{q}_3)$ is obtained from the PES reported by Li. et al [6].

The RPMD method introduces two dividing surfaces, defined in terms of the centroid coordinates of the ring polymers [22]. The first surface locates in the asymptotic reactant valley and is defined by the vector $\vec{R}$ connecting the centers of mass (COM) of both reactants (H and MgH)

$$s_0\left(\bar{q}\right) = R_\infty - \left|\bar{\vec{R}}\right|, \tag{2}$$

where $R_\infty$ is the distance where the interaction becomes negligible. The second dividing surface is placed in the vicinity of the transition state and it is defined in terms of the distance between the breaking and forming bonds [22]:

$$s_1\left(\bar{q}\right) = \left(\left|\bar{q}_{\mathrm{MgH}}\right| - q^{\ddagger}_{\mathrm{MgH}}\right) - \left(\left|\bar{q}_{\mathrm{HH}}\right| - q^{\ddagger}_{\mathrm{HH}}\right), \tag{3}$$

where $\bar{\vec{q}}_{\mathrm{AB}}$ is the vector between the centroids of the atoms A and B and $q^{\ddagger}_{\mathrm{AB}}$ is the corresponding distance of the atoms at the saddle point. With the two dividing surfaces defined above, the reaction coordinate $x$ can be written as

$$\xi\left(\bar{q}\right) = \frac{s_0\left(\bar{q}\right)}{s_0\left(\bar{q}\right) - s_1\left(\bar{q}\right)}. \tag{4}$$

Adapting the Bennett−Chandler factorization [23,24], the RPMD rate coefficient can be expressed by [20,25]

$$k_{\mathrm{RPMD}} = k_{\mathrm{QTST}}\left(T;\xi^{\ddagger}\right)\kappa\left(t\to\infty;\xi^{\ddagger}\right) f\left(T\right) \tag{5}$$

where the first term $k_{\mathrm{QTST}}\left(T;\xi^{\ddagger}\right)$ is the static contribution, denoted as the centroid-density quantum transition state theory (QTST) [26,27] rate coefficient [28], calculated from the peak position $\xi^{\ddagger}$ of the potential of mean force (PMF) curve along the reaction coordinate $\xi\left(\bar{q}_{\mathrm{AB}}\right)$ [24,30]:

$$k_{\mathrm{QTST}}\left(T;\xi^{\ddagger}\right)=4\pi R_{\infty}^{2}\left(\frac{1}{2\pi\beta\mu}\right)^{\frac{1}{2}}\mathrm{e}^{-\beta\left[W\left(\xi^{\ddagger}\right)-W(0)\right]},\tag{6}$$

where $\mu=\frac{m_{\mathrm{A}}m_{\mathrm{B}}}{m_{\mathrm{A}}+m_{\mathrm{B}}}$ is the reduced mass of both reactants, and $W\left(\xi^{\ddagger}\right)-W(0)$ is the difference of potential of mean force (PMF) from umbrella integration [22,29].

The second term $\kappa\left(t\rightarrow\infty;\xi^{\ddagger}\right)$ is the transmission coefficient representing the dynamic correction and accounting for the recrossing of the free-energy barrier $\xi^{\ddagger}$. This factor counterbalances $k_{\mathrm{QTST}}\left(T;\xi^{\ddagger}\right)$, ensuring the independence of the RPMD rate coefficient $k_{\mathrm{RPMD}}\left(T\right)$ of the selection of the dividing surface [20,21,25].

The third term $f(T)$ is the electronic degeneracy factor, and its value equals to the ratio between electronic partition functions of TS and reactants. Consider H is in $^{2}\mathrm{S}$ state, MgH is in $\mathrm{X}^{2}\Sigma^{+}$ state, and $MgH_2$ is in $^{1}\mathrm{A}'$ state [30], the value is:

$$f(T)=\frac{Q_{\mathrm{elec}}^{\mathrm{TS}}}{Q_{\mathrm{elec}}^{\mathrm{Reactant}}}=\frac{Q_{\mathrm{elec}}^{\mathrm{TS}}}{Q_{\mathrm{elec}}^{\mathrm{H}}\times Q_{\mathrm{elec}}^{\mathrm{MgH}}}=\frac{1}{2\times2}=\frac{1}{4}\tag{7}$$

When only one bead is used, RPMD rate coefficients reduce to the classical ones. And the minimal number of beads needed to fully account for the quantum effects can be estimated by the following formula [29]

$$n_{\min}=\beta\hbar\omega_{\max},\tag{8}$$

where $\omega_{\max}$ is the largest vibrational frequency of the reaction system.

**B. Obtaining the initial configurations for both reaction channels**

Since the title reaction contains two reaction channels, we developed a new protocol to obtain initial configurations for umbrella sampling. This protocol contains two steps and is called the optimization-interpolation method (OIM). Firstly, we start from TS and then use the steepest descent (SD) method to optimize the system downhill to both sides of the potential energy peak, constructing a set of initial configurations for each window for umbrella sampling. Such an optimization step is to ensure the generated configurations are in the desired minimum reaction path (MRP). Since the reaction coordinate for the optimized configuration $\{\bar{\boldsymbol{q}}_i^{'}\}_r^0, i = \mathrm{H_1, Mg, H_2}$ in each window $r$, $\xi_{r,0}^{'}$, would not be the same as the one wanted $\xi_{r,0}$, we perform the following interpolation step to obtain new configuration $\{\bar{\boldsymbol{q}}_i\}_r^0, i = \mathrm{H_1, Mg, H_2}$ corresponding to $\xi_{r,0}$. Fig. S1 in supporting information shows the two sets of initial configurations along the two reaction channels.

We also used an adaptive method to estimate the force constant for each window, $k_r$. The $k_r$ is set to $2.72\,(T/\mathrm{K})\,\mathrm{eV}$ at the beginning, and then we perform short trajectories within the window $\xi_{r,0}$, gradually increase $k_r$ in that window as $k_r^{\mathrm{new}} = 1.2 k_r^{\mathrm{old}}$, until the trajectory is properly constrained.

## III. COMPUTATIONAL DETAILS

We used the PES developed by Li et al [6]. The calculation parameters are collected in Table I. Especially, the thermal rate coefficients for each reaction channel were calculated at five temperatures ranging from 200 to 1500 K. Numbers of beads were used are 32 for $T \leq 300$ K, 16 for 500 K and 1000 K. The parameter $R_\infty$ in Eq. (2) was set to 12 Å for all of the temperatures considered.

As used in our previous work on H + $CH_4$ [31], we use different window sizes to calculate the PMF for both channels. In the relatively flat region of the entrance channel $(\xi < 0.65)$, wider windows are used. While in the transition-state region, smaller windows are chosen instead. In each window, the RPMD trajectory in the *NVT* ensemble was performed for 6 ns, which is tested converged. In calculating the transmission coefficient, for both channels, 40,000 unconstrained child trajectories were propagated for 10 ps with initial conditions sampled from a long parent trajectory with its centroid constrained at the transition-state geometry by SHAKE [32]. All calculations were performed with a time step of 0.1 fs. Finally, in all the trajectories, Anderson thermostat is used [33,34].

For comparison, we also used the canonical variational transition state theory with the microcanonically optimized multidimensional tunneling (CVT/μOMT) to calculate the rate coefficients [35-38]. The CVT/μOMT rate coefficients were calculated with version 2017-C of the POLYRATE program [39], and the variable INH is set to 100. In such calculations, we also calculated rate coefficients for both channels, and combine them to obtain the final results.

## IV. RESULTS AND DISCUSSION

We collect our results from RPMD calculations in Table II and will discuss the results in detail below.

Fig. 2 depicts profiles of the potential of mean forces (PMFs) along the reaction coordinate (left panel) and corresponding transmission coefficients (right panel) for the title reaction at 500 K in both channels. According to the previous work, the t-TS is after the complex well of H-Mg-H, as set as $x=1.0$. But according to our potential energy scan, we found before the complex well, there is a potential energy barrier at the entrance region; this barrier also provides a free energy barrier at $x\sim 0.6$. For the roaming channel, the reported potential energy is at the entrance before the H-Mg-H complex forming. But considering the flatness of the roaming channel, we assigned the dividing surface forward to the barrier, to ensure the free energy barrier can be correctly located. So in our reaction coordinate of roaming channel, the r-TS is at $x\sim 0.75$, and obtained free energy barrier is also near $x\sim 0.75$.

Since in both channels, there is a very deep potential well corresponding to the complex H-Mg-H after the TS region, we also show the detailed PMF curves at the TS region in subplots. From the plot, one can see the free energy barriers are very low and with similar hight. Such a flat free energy profile before the TS region can also explain why the title reaction has so much ratio of roaming. From the plot, we can also observe the heights of RPMD free energy barriers from the converged number of beads are slightly higher than that from 1-bead similar to that from some barrierless insertion reactions. As discussed before, increasing of free energy barrier would result from both the reduction of the entropy and the increase of the ZPE along the reaction coordinate [21].

The right panel of Fig. 2 shows the RPMD time-dependent transmission coefficients $\kappa(t)$ from both 1-bead and the converged number of beads and at 500 K. One can observe that the plateau time of $\kappa(t)$ from 1-bead is smaller than that from multi-bead. From previous RPMD studies [30,35], there are two general features of the transmission coefficients: (a) 1-bead transmission coefficients are always higher than corresponding multi-bead ones; (b) 1-bead transmission coefficients decrease with decreasing temperature while for multi-bead ones that trend is opposite. But it's not the case in the title reaction. Here we find the multi-bead transmission coefficients are larger than 1-bead ones, and that both 1-bead and multi-bead transmission coefficients decrease with increasing temperature, which can be also seen in Table I. This would stem from the fact that the title reaction is an addition at the first step, forming a complex H-Mg-H which forms a deep potential well, other than direct abstract reactions intensively investigated previously. Furthermore, in the insertion reactions X+$H_2$ (X=O, C, S, N) [21], we also observed the 1-bead and corresponding multi-bead transmission coefficients are close to each other, so the previously proposed trend would change in a different type of reactions.

Fig. 3 collects the converged RPMD free energy curves (left panel) and corresponding transmission coefficients (right panel) at different temperatures. It can be seen for the tight channel, the positions of all free-energy barriers $\xi^{\ddagger}$ are converging to 0.6 as the reaction temperature rises, while for the roaming transition state, the barrier positions are converging to 0.78. For both the two reaction channels, the free energy curve is monotonically increasing with temperature, due to the increase of thermal motion.

All the transmission coefficients $\kappa(t)$ decay quickly within 250 fs with frequent oscillations. But although at low temperatures (200 K and 300 K), the curves for both channels reach the plateau as fast as within 400 fs, at high temperatures (1000 K) the curves become hard to converge, reaching the plateau only after 1500 fs. This temperature dependence of converging time of $\kappa(t)$ is also different from observed in previous works. Such as in the insertion reactions [21], only at low temperature $\kappa(t)$ becomes slower to converge. The reason of the slowing convergence for $\kappa(t)$ maybe from the low barrier addition reaction feature on the PES, so when the temperature increases, the recrossing becomes heavy, but when the temperature decreases, the product state is easier to reach once the reaction system cross the barrier, and stay long in the deep potential well of complex H-Mg-H so that less recrossing occurs.

Fig. 4. with Table I also collects the thermal rate coefficients from RPMD, CVT/μOMT [35], and quantum dynamics (QD). For the case of RPMD, we also show those from single-channel (tight, $k_{\mathrm{RPMD}}^{\mathrm{tight}}$ and roaming, $k_{\mathrm{RPMD}}^{\mathrm{roaming}}$) and the total rate coefficient: $k_{\mathrm{RPMD}}^{\mathrm{total}} = k_{\mathrm{RPMD}}^{\mathrm{tight}} + k_{\mathrm{RPMD}}^{\mathrm{roaming}}$. But, we only depict the total rate coefficients from other methods.

At first glance, all the results are in the same order of magnitude, showing the generally temperature-independent feature roughly speaking and showing consistency among different methods. However, as one can also see, they exhibit different trends of chaining with temperature. The RPMD thermal rate coefficients decrease with increasing temperature from 300 K to 1000 K. At 200 K, the rate coefficient from RPMD becomes smaller than that at 300 K but still larger than that at 1000 K. This is because in the title reaction, the first step is a complex-forming H + MgH → H-Mg-H [40,41], and the trend mentioned above is typical for some of the low barrier addition reaction [42,43]. And the slight reduction of the rate coefficient at 200 K may come from the competition between the ZPE and tunneling. The CVT/μOMT rate coefficients also exhibit the similar trend, but with different line curvature. They also decrease with increasing temperature below 500 K. But when the temperature becomes higher, the results become similar to that from QD. And it's worth pointing out, when the temperature is as high as 1000 K, results from all three methods become much similar, reflecting the consistency at high-temperature limit. It's also worth pointing out here since the skew angle of the title reaction is about $46^{\circ}$, the recrossing correction in CVT/μOMT calculation will affect results less [44] so that the similarity between results from it and RPMD below 500 K would result from error cancellation. And one can also see the 1-bead RPMD rate coefficients are lower than converged RPMD ones at lower temperatures. This show again RPMD includes quantum dynamic features which absent in classical MD. Moreover, our RPMD results are with opposite trend with temperature compared with the QD rate coefficient [40]. That may result from the fact that those QD results from calculations on the ground vibrational and rotational state ($v=0, j=0$) only, the vibrationally excited states play a considerable role in the quantum dynamical thermal rate coefficient.

## V. CONCLUSION

In this work, we have used RPMD to calculate the thermal rate coefficients of a roaming reaction H + MgH → Mg + $H_2$ in the temperature range of 200 to 1000 K. The results have been compared with a version of VTST method CVT/μOMT and the previous theoretical QD work.

By exerting a two-step optimization-interpolation method, we successfully generated the initial configurations of the system in both the two reaction channels and calculated RPMD rate coefficients of each channel. By combining them, we obtain the total thermal rate coefficients for the title reaction. To our knowledge, this is the first time that a multi-channel reaction has been studied using RPMD.

And we also find both the RPMD and CVT/μOMT rate coefficients decrease with the increasing temperature. This finding is different from previous QD results but can be understood as a typical feature of addition reaction since the first step of title reaction is a complex-forming. We hope experimental work or accurate QD calculations in the future can confirm that.

## VI. ACKNOWLEDGEMENTS

This study was funded by the National Nature Science Foundation of China (No. 21503130 and 11674212 to Y.L., and 21603144 to J.S.). Y.L. is also supported by the Young Eastern Scholar Program of the Shanghai Municipal Education Commission (QD2016021) and the Shanghai Key Laboratory of High Temperature Superconductors (No. 14DZ2260700). J.S. is also supported by Shanghai Sailing Program (No. 2016YF1408400). Part of the calculations has been done on the supercomputing system, Ziqiang 4000, in the High Performance Computing Center at Shanghai University. And Y.L. thanks Bin Jiang and Anyang Li for their helpful discussions.

Figure 1. Two-dimensional contour plot of the PES in Jacobi coordinates with the Mg-H distance fixed at its equilibrium $r = 1.82$ Å. Both the tight and roaming transition states are indicated (red dots). The solid contours have an interval of 5 kcal/mol, and the dashed contours have an interval of 0.05 kcal/mol from −1 to 0 kcal/mol. The black dot line represents the configurations of each window and the red dot corresponds to TS. The barrier heights of both TS are shown too.

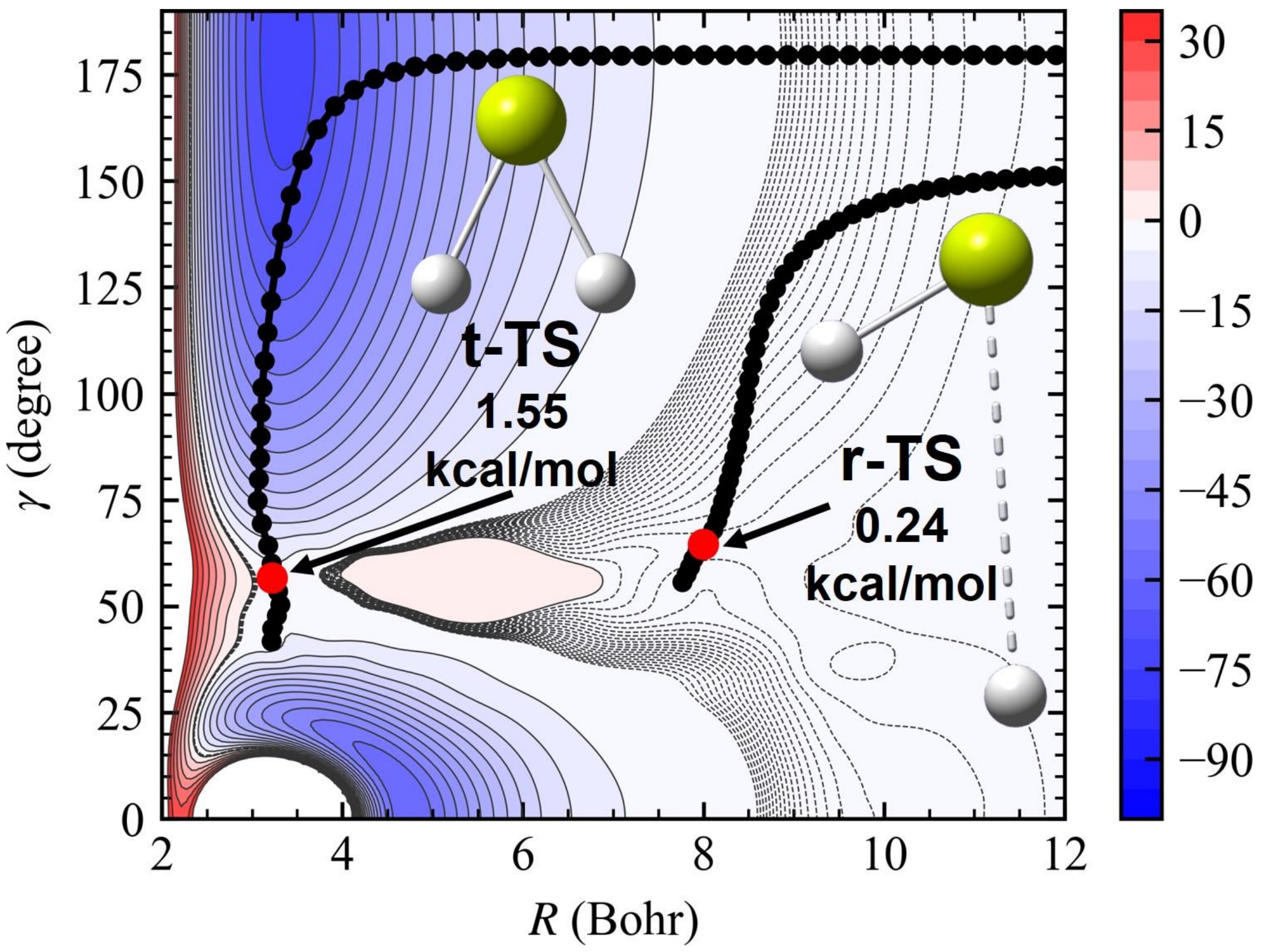

Figure 2. Comparison of the potential of mean force (PMF) (left panels) and transmission coefficients (right panels) for the H + Mg reaction at 500 K with 1-bead and the converged number of beads.

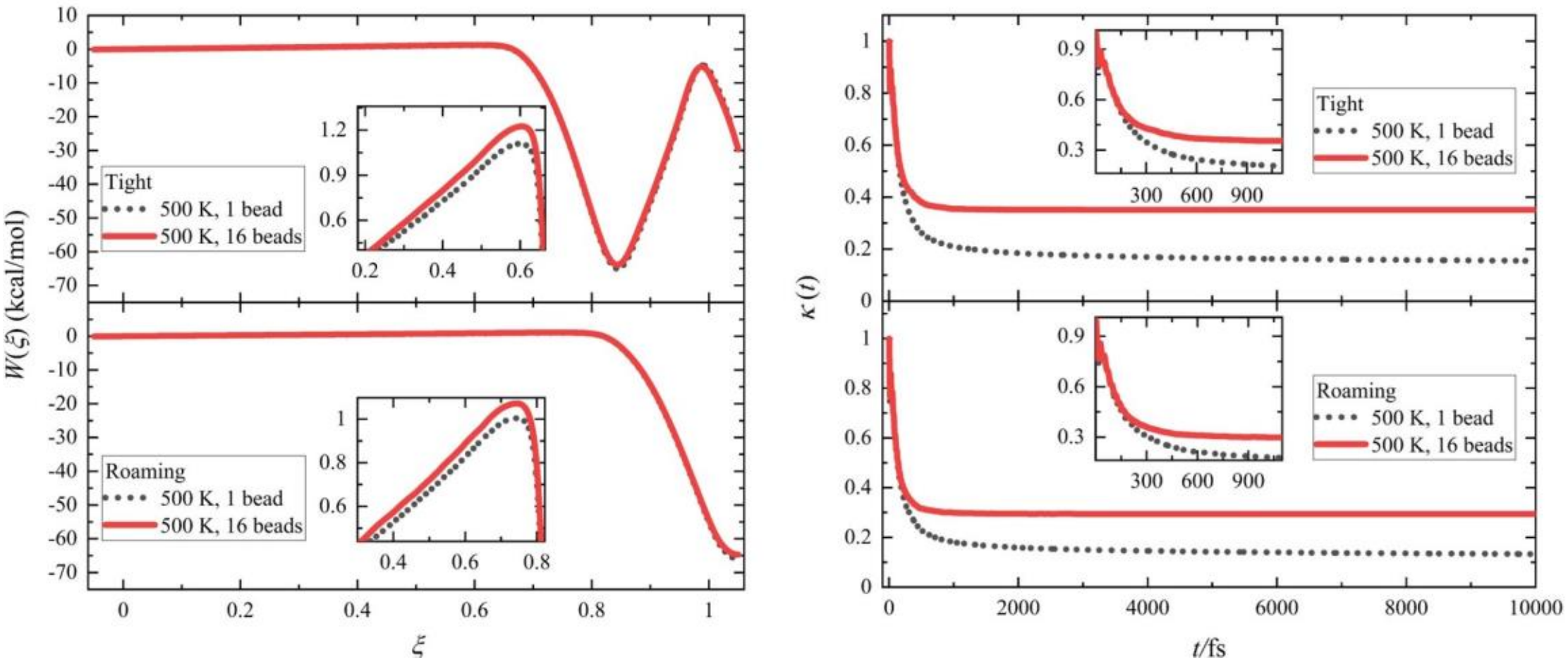

Figure 3. Converged ring-polymer PMF (left panel) and $\kappa(t \to \infty)$ (right panel) along the reaction coordinate for the H + MgH chemical reaction at 200, 300, 500, and 1000 K.

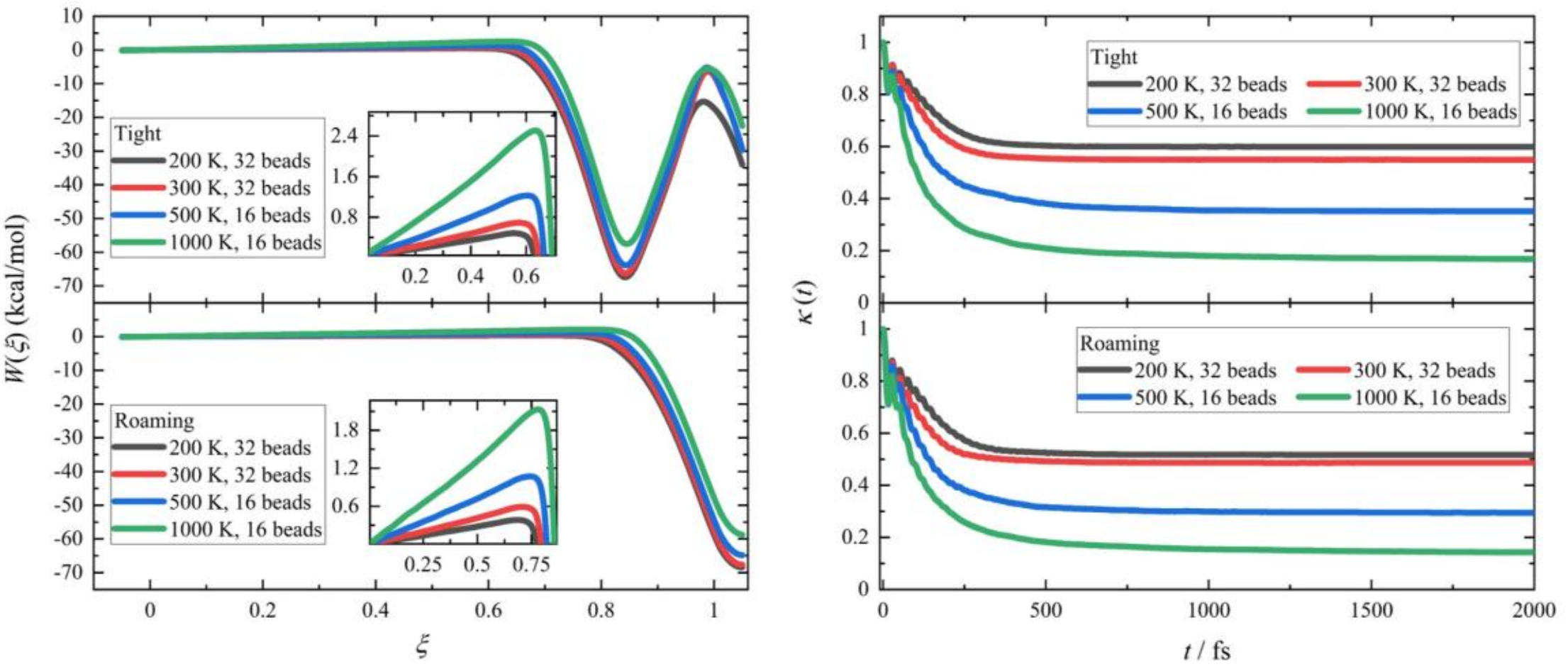

Figure 4. Arrhenius plot of RPMD (total: red, tight channel: green and roaming channel: blue), CVT/μOMT (orange), and QD (black) rate coefficients for the H + MgH reaction between 200 K and 1000 K.

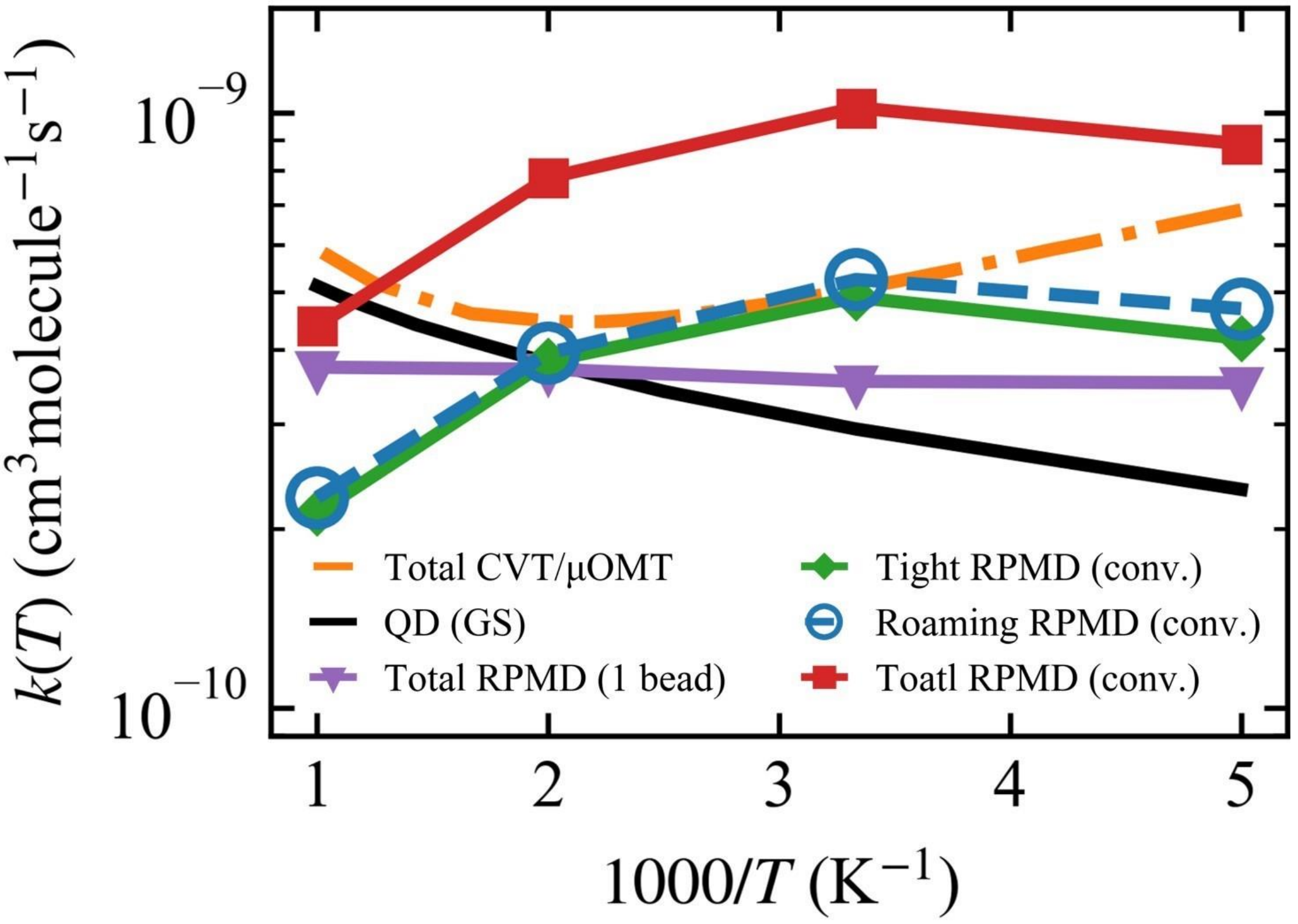

TABLE I. Input parameters for the RPMD calculations on title reactions.

| Parameter | Reaction | | Explanation |
|---|---|---|---|
| H + MgH | | | |
| Command line parameters | | | |
| | 200 | 500 | |
| Temp | 300 | 1000 | Temperature (K) |
| $N_{\text{beads}}$ | 32 | 16 | Number of beads |
| Dividing surface parameters | | | |
| $R_{\infty}$ | 12 | | Dividing surface $s_1$ parameter ($a_0$) |
| $N_{\text{bonds}}$ | 1 | | Number of forming and breaking bonds |
| $N_{\text{channel}}$ | 1 | | Number of equivalent product channels |
| Thermostat | Andersen | | Thermostat option |
| Biased sampling parameters | | | |
| $N_{\text{windows}}$ | 111 | | Number of windows |
| $\xi_1$ | −0.05 | | Center of the first window |
| d$\xi$ | 0.01 | | Window spacing step |
| $\xi_{\text{N}}$ | 1.05 | | Center of the last window |
| d$t$ | 0.0001 | | Time step (ps) |
| $N_{\text{trajectory}}$ | 6000 | | Number of trajectories |
| $t_{\text{equilibration}}$ | 0.2 | | Equilibration period (ps) |
| $t_{\text{sampling}}$ | 1 | | Sampling period in each trajectory (ps) |
| Potential of mean force calculation | | | |
| $\xi_0$ | −0.05 | | Start of umbrella integration |
| $\xi_{\ddagger}$ | 1.05 | | End of umbrella integration |
| $N_{\text{bins}}$ | 10,000 | | Number of bins |
| Recrossing factor calculation | | | |
| d$t$ | 0.0001 | | Time step (ps) |
| $t_{\text{equilibration}}$ | 20 | | Equilibration period (ps) in the constrained (parent) Trajectory |
| $N_{\text{totalchild}}$ | 40,000 | | Total number of unconstrained (child) trajectories |
| $t_{\text{childsampling}}$ | 10 | | Sampling increment along the parent trajectory (ps) |
| $N_{\text{child}}$ | 500 | | Number of child trajectories per one initially constrained Configuration |
| $t_{\text{child}}$ | 10 | | Length of child trajectories (ps) |

TABLE II. Compare the RPMD rate coefficients (unit $cm^3$ $molecule^{-1}s^{-1}$) of the H + MgH reaction with other published theoretical results.

| T/K | 200 | 300 | 500 | 1000 |
|---|---|---|---|---|
| Tight channel | | | | |
| $f(T)$ | 0.25 | | | |
| $N_{\text{beads}}$ | 32 | 32 | 16 | 16 |
| $\xi^{\ddagger}$ | 0.558 | 0.572 | 0.600 | 0.632 |
| $\Delta G(\xi^{\ddagger})$ | 0.48 | 0.69 | 1.22 | 2.49 |
| $k_{\text{QTST}}$ | $2.80\times10^{-9}$ | $3.64\times10^{-9}$ | $4.36\times10^{-9}$ | $6.03\times10^{-9}$ |
| $\kappa(t\to\infty)$ | 0.598 | 0.538 | 0.355 | 0.141 |
| $k_{\text{RPMD}}^{\text{tight}}$ | $4.18\times10^{-10}$ | $4.90\times10^{-10}$ | $3.83\times10^{-10}$ | $2.13\times10^{-10}$ |
| $k_{\text{CVT}/\mu\text{OMT}}^{\text{tight}}$ | $5.68\times10^{-10}$ | $3.34\times10^{-10}$ | $1.73\times10^{-10}$ | $7.42\times10^{-11}$ |
| Roaming channel | | | | |
| $f(T)$ | 0.25 | | | |
| $N_{\text{beads}}$ | 32 | 32 | 16 | 16 |
| $\xi^{\ddagger}$ | 0.681 | 0.709 | 0.736 | 0.772 |
| $\Delta G(\xi^{\ddagger})$ | 0.38 | 0.59 | 1.03 | 2.13 |
| $k_{\text{QTST}}$ | $3.59\times10^{-9}$ | $4.29\times10^{-9}$ | $5.32\times10^{-9}$ | $7.25\times10^{-9}$ |
| $\kappa(t\to\infty)$ | 0.520 | 0.486 | 0.296 | 0.124 |
| $k_{\text{RPMD}}^{\text{roming}}$ | $4.69\times10^{-10}$ | $5.25\times10^{-10}$ | $3.95\times10^{-10}$ | $2.26\times10^{-10}$ |
| $k_{\text{CVT}/\mu\text{OMT}}^{\text{roaming}}$ | $1.19\times10^{-10}$ | $1.72\times10^{-10}$ | $2.74\times10^{-9}$ | $5.15\times10^{-9}$ |
| $k_{\text{RPMD}}^{\text{total}}$ [a] | $8.87\times10^{-10}$ | $1.02\times10^{-09}$ | $7.78\times10^{-10}$ | $4.39\times10^{-10}$ |
| $k_{\text{CVT}/\mu\text{OMT}}^{\text{total}}$ [b] | $6.87\times10^{-10}$ | $5.06\times10^{-10}$ | $4.47\times10^{-10}$ | $5.90\times10^{-10}$ |
| $k_{\text{QD}}$ [c] | $2.33\times10^{-10}$ | $2.94\times10^{-10}$ | $3.80\times10^{-10}$ | $5.12\times10^{-10}$ |

a: $k_{\text{RPMD}}^{\text{total}} = k_{\text{RPMD}}^{\text{tight}} + k_{\text{RPMD}}^{\text{roaming}}$

b: $k_{\text{CVT}/\mu\text{OMT}}^{\text{total}} = k_{\text{CVT}/\mu\text{OMT}}^{\text{tight}} + k_{\text{CVT}/\mu\text{OMT}}^{\text{roaming}}$

c: This is the total reaction rate coefficient from QD calculations.

**TOC**

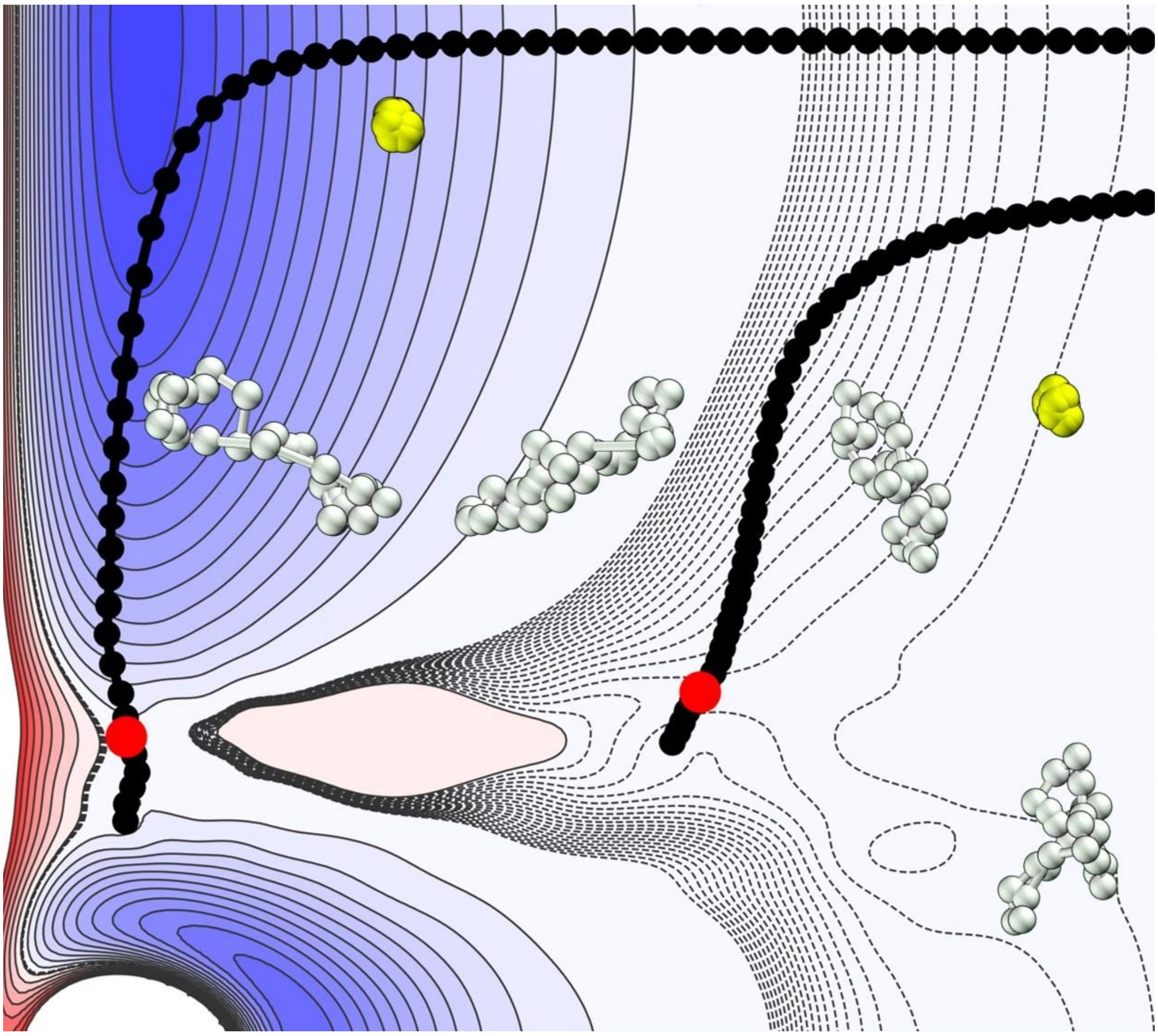

Supporting information

Figure S1. The minimum reaction path (MRP) along with the reaction coordinates. The black dotted line is the minimum energy path after optimization, and the red dotted line is the minimum energy path after spline interpolation.

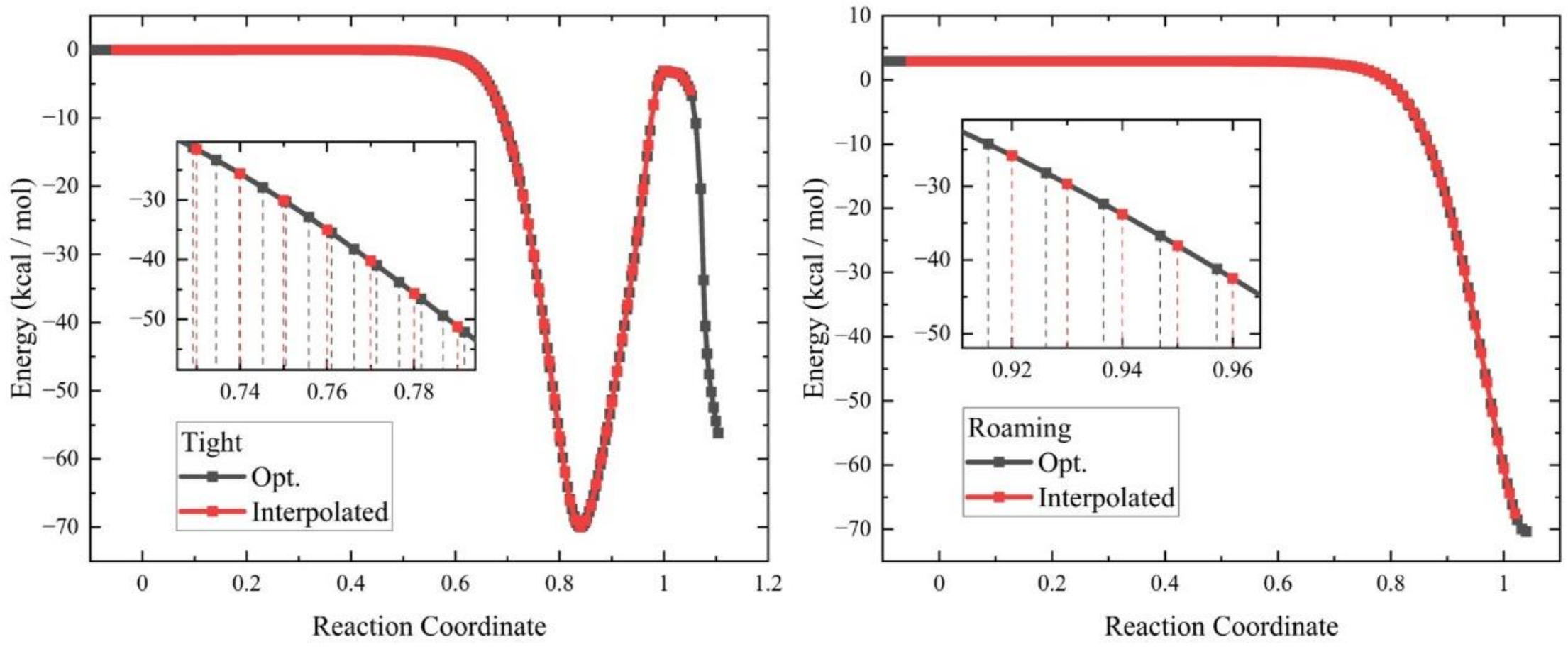